\documentclass{elsart}

\usepackage{graphicx}
\usepackage{amssymb}
\usepackage{epsfig}
\usepackage{subfigure}
\usepackage{color}
\usepackage[english]{babel}
\usepackage{latexsym}
\usepackage{amsfonts}
\usepackage{amsmath}
\usepackage{multirow}
\usepackage{float}
\usepackage{longtable}

\def\be{\begin{equation}}
\def\ee{\end{equation}}

\begin{document}
\begin{frontmatter}

\title{Thermal description of hadron production in $e^{+}e^{-}$ collisions revisited}

\author[gsi]{A.~Andronic},
\author[hei]{F.~Beutler}
\author[gsi,emmi,tud]{P.~Braun-Munzinger},
\author[wro,tud]{K.~Redlich},
\author[hei]{J.~Stachel}

\address[gsi]{GSI Helmholtzzentrum f\"ur Schwerionenforschung
D-64291 Darmstadt, Germany}
\address[emmi]{ExtreMe Matter Institute EMMI, GSI, D-64291 Darmstadt, Germany}
\address[hei]{Physikalisches Institut der Universit\"at Heidelberg,
D-69120 Heidelberg, Germany}
\address[tud]{Technical University Darmstadt, D-64289 Darmstadt, Germany}
\address[wro]{Institute of Theoretical Physics, University of Wroc\l aw,
PL-50204 Wroc\l aw, Poland}

\begin{abstract}
  We present a comprehensive analysis of hadron production in $e^{+}e^{-}$
  collisions at different center-of-mass energies in the framework of the
  statistical model of the hadron resonance gas.  The model is formulated in
  the canonical ensemble with exact conservation of all relevant quantum
  numbers. The parameters of the underlying model were determined using a fit 
  to the average multiplicities of the latest measurements at 
  $\sqrt{s}$ = 10, 29-35, 91 and 130-200 GeV. 
  The results demonstrate that, within the accuracy of the
  experiments,  none of the data sets is satisfactorily described with
  this approach, calling into question the notion that particle production in
  $e^{+}e^{-}$ collisions is thermal in origin.
\end{abstract}

\end{frontmatter}

\section{Introduction}

The analysis of hadron yields measured in central heavy ion collisions from
AGS up to RHIC energies has shown
\cite{agssps,satz,heppe,cley,beca1,rhic,nu,beca2,rapp,becgaz,aat} that hadron
multiplicities can be described very well with a hadro-chemical equilibrium
approach which is governed by the chemical freeze-out temperature T,
baryo-chemical potential $\mu_b$, and the fireball volume V; for a recent
review see \cite{review}.  The natural question arising here is whether this
statistical behavior is a unique feature of high energy nucleus-nucleus
collisions or whether it is also applicable in elementary collisions like,
e.g., $e^{+}e^{-}$.  Previous publications indicated
that indeed hadron production in $e^{+}e^{-}$ collisions at 14-43 GeV
\cite{Becattini:1995if,Becattini:1996gy,becattini01} and 91 GeV
\cite{Becattini:1995if,Becattini:1997uf} can be well described within a
thermal model provided that local quantum number conservation is properly
implemented.  The main result of these investigations was that the temperature
values deduced are almost constant near T = 160 MeV and that the volume
increases with energy, while strangeness is undersaturated. These results were
taken, together with the results for nucleus-nucleus collisions where a
similar temperature is reached at high energies, as evidence for the
interpretation that the thermodynamical state is not reached by dynamical
equilibration among constituents but rather is a generic fingerprint of
hadronization \cite{stock,heinz} or a feature of the excited QCD vacuum
\cite{castorina}. 
Alternatively, it was argued in \cite{wetterich} that the quark-hadron 
phase transition drives the equilibration dynamically for nucleus-nucleus 
collisions. Equilibration in  $e^{+}e^{-}$ collisions is not 
easy to explain in this latter approach.   

In our new analysis we are using the latest multiplicity measurements
summarized and published by the Particle Data Group (PDG) \cite{pdg}.
Since the aim is a precision calculation for a small system we employ a fully
canonical form of the statistical model \cite{review,redlich} conserving 
baryon number N, charge Q, strangeness S, charmness C, and bottomness B. 
In the present analysis charmed and bottom hadrons are relevant only via 
their feed-down contributions to the yields of the lighter hadron species
(for more details see below).  
To reach within the model a precision comparable to that of the data from 
the LEP collider (a few percent), we have performed computations 
including quantum statistics (see below).

\section{The model}

The canonical statistical model we will base our investigations on  is described in
\cite{review,Becattini:1995if,Becattini:1996gy,Becattini:1997uf,redlich}.
Here we present a short summary with emphasis on the way the quantum number
conservation is implemented.
Most hadronic events in high energy e$^+$e$^-$ annihilations are two-jet events,
originating from quark-antiquark pairs of the five lightest flavors.
Since we would like to address the issue of overall equilibration in these systems,
we are using a scheme in which each jet is treated as a fireball with vanishing 
quantum numbers as fixed by the entrance channel; we call this the ``no flavor'' 
scheme. It is clear at this point that hadrons from jets with heavy quarks 
($c$ and $b$) will be greatly underestimated by the model because of the large 
Boltzmann suppression factors.
In this approach the issue of equilibration is effectively addressed only for 
hadrons with light quarks ($u$, $d$, $s$).

It is important to recognize that the measured yields of these hadrons contain 
the contribution from the e$^+$e$^-$ annihilation events into $c\bar{c}$ and 
$b\bar{b}$.
Heavy-quark production is indeed significant and is very precisely measured,
in particular at the $Z^0$ mass ($\sqrt{s}$=91.2 GeV), where the measurements 
are very well described by the standard model \cite{pdg}.
Hence, heavy-quark production is manifestly non-thermal in origin. 
We therefore consider three scenarios: 
i) we fit the data as measured.
ii) we subtract from the yields of hadrons carrying light quarks 
the contribution originating from charm and bottom decays based on available 
data for charmed and bottom hadron production (and their branching ratios)
at 91 GeV.
iii) we perform a fit to the data at 91 GeV in a 5-flavor ($u$, $d$, $s$, $c$,
$b$) approach.
In this case e$^+$e$^-\rightarrow q\bar{q}$ events are treated in as
2-jet initial state where each jet carries the relevant quantum numbers. 
The fractions of the quark flavors in hadronic events \cite{pdg} are thus external 
input values, unrelated with the thermal model (see also Table II in 
ref.~\cite{becattini08}).
We will treat the heavy quark sector in detail in a forthcoming
publication \cite{xyz}. 

Scenarios i) and ii) differ from that used e.g. in ref. 
\cite{becattini08}, where a 5-flavor scheme is used throughout.
A significant difference between our approach and that of
\cite{becattini08} is in the treatment of the volume entering the statistical
model calculations, as explained below. In our case the volume is
the hadronization volume of each jet, meaning that each jet hadronizes
separately, as in deep inelastic scattering. The yields calculated for each
jet are then added to compare with data. In ref. \cite{becattini08} the volume 
is that of both jets together. Because of different canonical corrections 
(see below) this leads to significant differences in the final results. 
While we consider our jet hadronization picture
as more suited to describe two-jet events within a statistical 
framework\footnote{Three and more jet events are neglected here as they are 
in other approaches too.}, 
we acknowledge that the approach of hadronization in one volume 
\cite{becattini08} cannot be excluded at present.

The appropriate tool to deal in a statistical mechanics framework with a system 
where all quantum numbers are zero is the canonical partition function with exact 
conservation of the N, Q, S, C and B quantum numbers \cite{turko81}:
\begin{equation}\label{eq:partition1}
\begin{split}
Z(\vec{X}) &= \frac{1}{(2\pi)^5}\int d^5\vec{\phi}\; exp\{\sum^{N_B}_{j=1}\sum_k 
\ln(1-e^{-\beta \epsilon_{j,k}-i\vec{x}_j\vec{\phi}})^{-1}\\
&\quad+ \sum^{N_F}_{j=1}\sum_k \ln(1 + e^{-\beta \epsilon_{j,k}-i\vec{x}_j\vec{\phi}})\}
\end{split}
\end{equation}
where $\vec{x}_j$ is a five component vector $\vec{x}_j = (N_j, Q_j, S_j, C_j, B_j)$ 
containing the quantum numbers of hadron $j$, and 
$\phi = (\phi_N, \phi_Q, \phi_S, \phi_C, \phi_B)$ contains the parameters of the 
symmetry group $[U(1)]^5$. 
The sum over $k$ is for the phase space cells for each hadron, $\beta=1/T$, 
$\epsilon_{j,k}=\sqrt{p^2_k+m^2_j}$.
In this expression, each $\phi_X$ corresponds to the conservation of the 
corresponding quantum number $X$; $N_B$ and $N_F$ count the total number of boson 
and fermion states, respectively.

The usual procedure is to solve Eq.~\ref{eq:partition1} in the Boltzmann 
approximation, i.e. $\ln(1\pm x)^{\pm 1}\approx x$
To perform the calculation with full quantum statistics we employ the series 
expansion\footnote{We perform calculations with quantum statistics for bosons 
only. Due to the higher masses, for fermions the Boltzmann approximation is 
sufficiently accurate.}
\begin{equation}
\ln(1-x)^{-1} = \sum^{\infty}_{k=1}\frac{x^{k}}{k}
\label{eq:partition19.2}
\end{equation} 
in Eq.~\ref{eq:partition1} leading to
\begin{equation}
Z(\vec{X}) = \frac{1}{(2\pi)^5}\int d^5\vec{\phi}\; e^{i\vec{X}\vec{\phi}} 
\exp\{\sum_j z^1_je^{-i\vec{x}_j\vec{\phi}} + 
\sum_{b}\sum_{k=2}^{\infty}z^k_be^{-ik\vec{x}_b\vec{\phi}}\}
\label{eq:partition2}
\end{equation}
with
\begin{equation}
z^k_j = g_j\frac{V}{k(2\pi)^3}\int d^3p\; e^{-\frac{\sqrt{\vec{p}^2+m_j^2}}{T}k}
\label{eq:partition3}
\end{equation}
where $m_j$ is the particle mass and $g_j$ is the spin-isospin degeneracy factor. 
The index $j$ runs over all particles species in the hadronic gas and $b$ runs 
only over bosons. Here and in the following we use units with $\hbar=c=1$.
The complete derivation of the partition function (for quantum statistics) 
will be the subject of a separate publication \cite{beutler_diploma}. 
The differences between calculations with Boltzmann and quantum statistics are 
presented in Table~\ref{tabl:comparison}.  

The set of hadron species used here was updated compared to \cite{aat} with
the most recent information available from the PDG \cite{pdg}
and includes all resonances listed there\footnote{We have included the $\sigma$
meson, as in ref.~\cite{aas}}. 
We have also, as in ref. \cite{aat}, included the resonance width 
explicitely\footnote{We have not used the method applied in 
\cite{Becattini:1995if,Becattini:1996gy,becattini01,Becattini:1997uf}
to include the uncertainties in hadron mass, width and branching ratios 
as additional systematic errors.}.

The integral representation of the partition function in
Eq.~(\ref{eq:partition2}) is not convenient for numerical analysis as the
integrand is a strongly oscillating function. Thus we have applied an
expansion into series of Bessel functions \cite{cle91,Cleymans:1997ib,pbm4} 
to obtain a result that is free of oscillations. We obtain the multiplicity 
$\langle n_j \rangle$ per jet for  particle species $j$ by introducing a 
fugacity parameter $\lambda_j$ which multiplies the particle partition 
function $z^k_j$ and by differentiating
\begin{equation}
\left.\langle n_j\rangle = \frac{\partial\ln Z}{\partial
\lambda_j}\right|_{\lambda_j=1}.
\label{eq:diff} 
\end{equation} 
One has to take into account that $\langle n_j \rangle$  is  the
yield resulting from one jet. The 
multiplicity for the whole event is then the sum over the two jets.

\begin{table}[htb]
\caption{Comparison of particle yields obtained with our code and with the THERMUS 
code \cite{Wheaton:2004qb} for both initial production (prior to strong decays)
and for the final values (after strong and electromagnetic decays). 
We show yields (sum of particles and antiparticle yields and for the 2 jets) 
calculated with our code with quantum statistics (QS) and with 
Boltzmann  statistics (BS) for the  parameter set T=158 MeV, V=30 fm$^3$ 
and $\gamma_s$=0.80.}
\label{tabl:comparison}

\vspace{.2cm}
\begin{tabular}{|c|c|c|c|c|c|c|}
\hline
& \multicolumn{4}{|c|}{this work} & \multicolumn{2}{|c|}{THERMUS}\\
\cline{2-7}
particle & \multicolumn{2}{|c|}{final} & \multicolumn{2}{|c|}{initial} & 
initial & final\\
\cline{2-5}
 & QS & BS & QS & BS & & \\ \hline
$\pi^{+}$ & 18.54 & 18.07 & 5.12 & 4.74 & 4.72 & 14.28 \\
$\pi^0$  & 11.08 & 10.73 & 2.97 & 2.68 & 2.68 & 8.29 \\
K$^+$ & 2.028 & 2.016 & 0.954 & 0.945  & 0.940 & 1.89 \\
K$^0$ & 1.952 & 1.939 & 0.938 & 0.928 & 0.924 & 1.84 \\
$\eta$ & 1.090 & 1.087 & 0.472 & 0.468 & 0.472 & 0.890 \\
$\rho^0$(770) & 1.12 & 1.12 & 0.756 & 0.753 & 0.756 & 1.044 \\
K$^{*0}$(892) & 0.597 & 0.595 & 0.447 & 0.445 & 0.442 & 0.570 \\
p           & 0.974 & 0.966 & 0.232 & 0.230 & 0.229 & 0.668 \\
$\phi$(1020) & 0.131 & 0.131 & 0.128 & 0.128 & 0.128 & 0.132 \\
$\Lambda$ & 0.364 & 0.361 & 0.0706 & 0.0703  & 0.0677 & 0.239 \\
$\Sigma^{+}$(1385) & 0.0393 & 0.0389 & 0.0313 & 0.0310 & 0.0300 & 0.0316 \\
$\Xi^-$ & 0.0202 & 0.0200 & 0.0115 & 0.0114  & 0.0108 & 0.0190 \\
$\Xi^0$(1530) & 0.00768 & 0.00760 & 0.00738  & 0.00731 & 0.00679 & 0.00679 \\
$\Omega$ & 0.00125 & 0.00123 & 0.00125 & 0.00123 & 0.00117 & 0.00117 \\
\hline
\end{tabular}
\end{table}

Before engaging in the present data analysis we have performed a comparison of
results from the model described above with those obtained using the THERMUS 
code \cite{Wheaton:2004qb}, which is available publicly. The results for a
particular set of parameters are presented in Table~\ref{tabl:comparison}. 
Here and in the following the quoted yields include the corresponding antiparticle 
states (i.e. the yield labelled $\pi^+$ is actually the sum of the yields of 
$\pi^+$ and $\pi^-$) and are the sum over the two jets, as summarized and 
published by the PDG \cite{pdg}.
The agreement between the particle yields obtained with both codes is
generally very good, lending strong support also to the numerical implementation 
of our methods.
Note that, in our calculations, we employ quantum statistics, whereas in the 
THERMUS code the Boltzmann approximation is used for the canonical ensemble. 
Neglecting quantum statistics causes an error of about 3\% for the final pion 
yield, i.e. larger than the uncertainty in the data.  
As expected, the effect on all other hadron yields is smaller. 
If we employ the Boltzmann approximation in our code we get, prior to strong
decays, values for mesons and non-strange baryons which are in agreement with 
THERMUS results at the percent level, while for strange baryons our yields
are systematically higher by about 2\% compared to those obtained with THERMUS.
Inspecting the yields after strong decays, one notices a discrepancy
between results from our code and from THERMUS for most of the hadrons.
Our larger yields are due to a more complete set of hadron species used in the 
calculations. Concerning the decays, we have ensured that, in our code, the 
decay tables are symmetrical for particles and antiparticles and the decay 
widths always add up to the total width, even if particular channels are 
not measured. As the branching ratios are not well known for some of the 
high-mass states, we have used the known BR of the nearest state with the same 
quantum numbers.

\section{The fit procedure} 

The multiplicity calculation proceeds in two steps. First, 
a primary hadron yield, $N_h^{th}$, is calculated  using
(\ref{eq:partition1}) and (\ref{eq:diff}). A crucial assumption of the
model is that the final yields of all particles are fixed at a common
temperature, the chemical decoupling point. As a second step all resonances
in the gas which are unstable against strong and electromagnetic decays are 
allowed to decay into lighter stable hadrons, using appropriate branching 
ratios ($B$) and multiplicities ($M$) for the decay $j \rightarrow h$ published 
by the PDG \cite{pdg}. The abundances in the final state are thus 
determined by
\begin{equation} 
N_h = N_h^{th} + \sum_j N_j \cdot B(j\rightarrow h) M(j\rightarrow h) 
\end{equation} 
where the sum runs over all hadron species.

In the resonance gas model the results are determined by the basic thermal 
parameters, temperature T and volume V (the volume corresponding to one jet).  
Following the approach of ref.\cite{Becattini:1995if},
we also introduce an additional parameter $\gamma_s$ into the  partition
function to account for a possible deviation of
strange particle yields from their chemical equilibrium values. 
If a hadron contains $n_s$ strange valence quarks, its production is reduced 
by a factor $\gamma_s^{n_s}$. This parameter is also applied to neutral mesons 
such as $\eta, \eta', \phi, \omega, f_2(1270)$ and $f_2'(1525)$ according to 
the fraction of $s\overline{s}$ content in the meson itself. The relevant 
fraction is determined using mixing formulae quoted in \cite{pdg}.

For the fit procedure we use the complete set of all measured yields of
hadrons carrying light quarks. 
The second scenario, namely subtracting the contribution originating from charm
and bottom decays, is investigated only at $\sqrt{s}$=91 GeV as a case study,
since the measurements needed to allow the subtraction are complete only at 
this energy. The $\chi^2$ fit is performed by minimizing  
\begin{equation} \chi^2 =
\sum_h\frac{(N_h^{exp}-N_h)^2}{\sigma_h^2} 
\end{equation} 
as a function of the three parameters T, V and $\gamma_s$, taking account of
the experimental uncertainties $\sigma_h$. 

\section{Results}

\begin{figure}[htb]
\centering\includegraphics[width=0.8\textwidth]{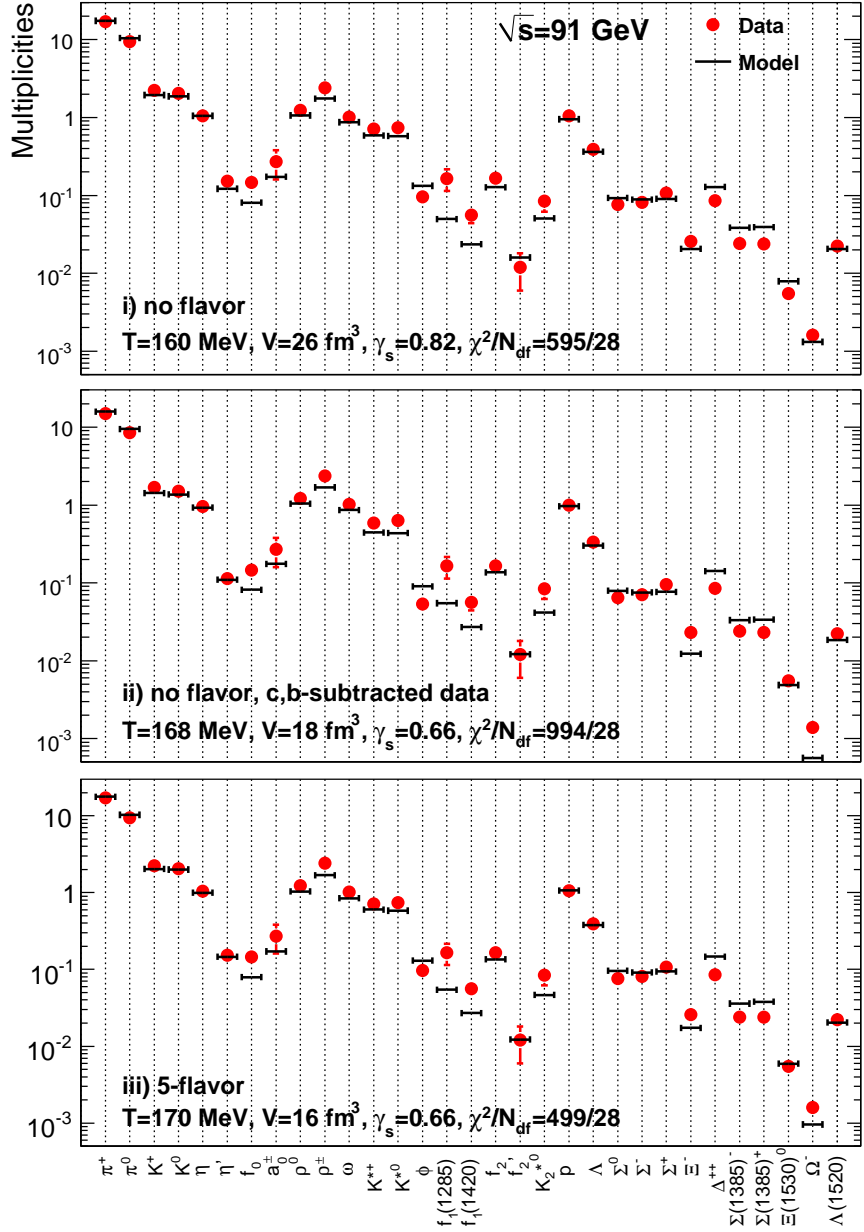}
\caption{Comparison between the best fit thermal model calculations and 
experimental hadron multiplicities (sum of particles and antiparticle yields 
and for the 2 jets) for $e^+e^-$ collisions at $\sqrt{s}$=91 GeV. 
The upper panel shows the fit in the non-flavor scheme of data including 
the feed-down contribution from heavy quarks, the middle panel is after subtraction 
of this contribution, the lower panel is for a 5-flavor scheme fit where the
flavor abundancies are extra input parameters from data (see text). 
The best fit parameters are listed for each case.}  
\label{fig:91GeV}
\end{figure}

The resulting best fit to the data at the energy $\sqrt{s}$=91 GeV  is shown 
in Fig.~\ref{fig:91GeV}. 
Shown are the three cases of the fit, the non-flavor approach without and with 
the subtraction of the contribution from heavy quarks (see below for the magnitude 
of this contribution) and the 5-flavor approach.
We first note the overall behavior of the data, namely an approximately
exponential decrease of particle yield with increasing particle mass. Such a
behavior is expected  in the hadron resonance gas model due to the  Boltzmann
factors, thus indicating the presence of statistical features of hadron production
in elementary collisions.

\begin{figure}[htb]
\begin{center}
\includegraphics[width=.93\textwidth]{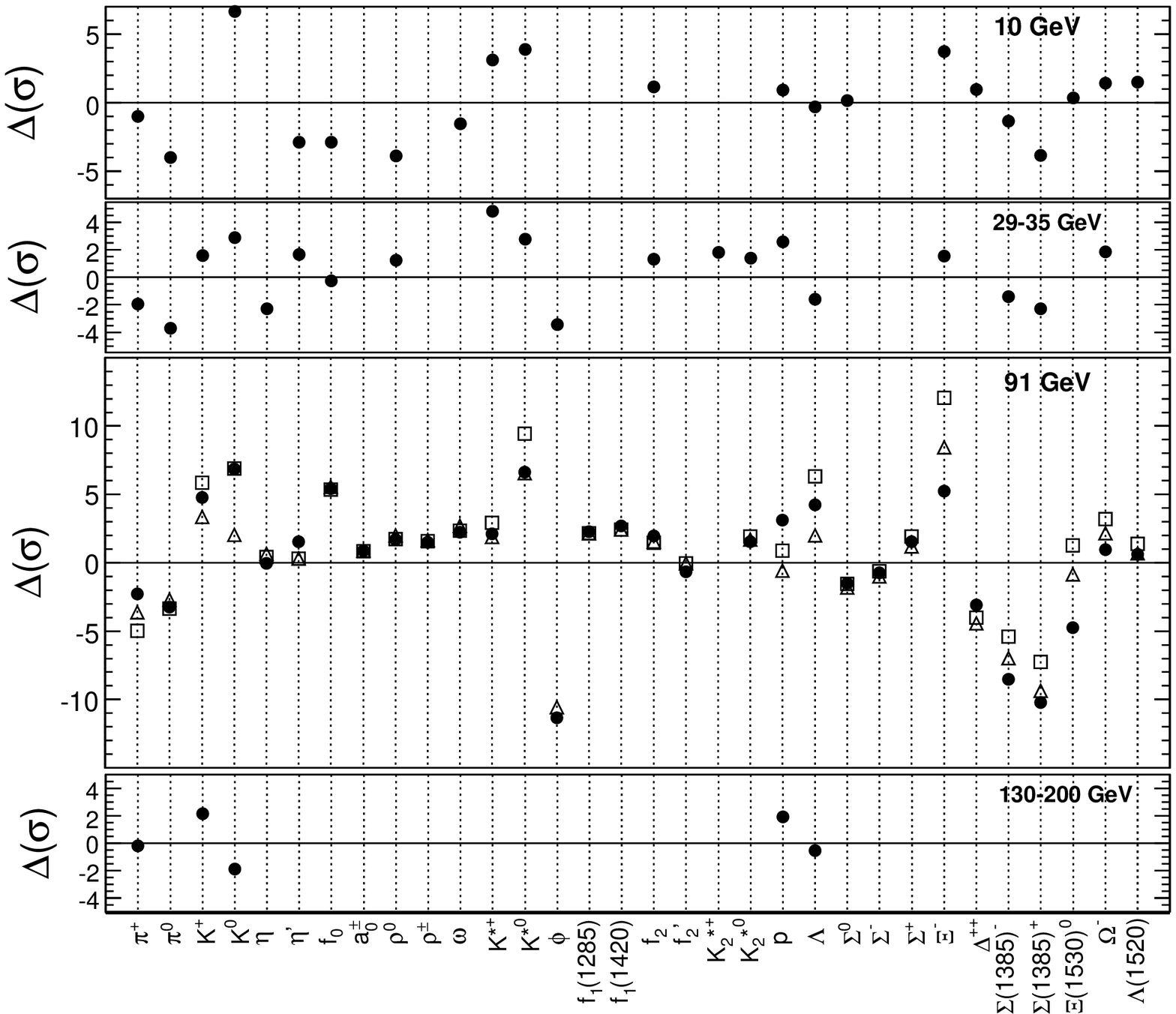}
\caption{Difference (in units of experimental error) between experimental data and
thermal model fits at four energies. 
For $\sqrt{s}$=91 GeV with open squares we show the results of the fit to data
after subtraction of the heavy quark contribution (scenario ii),
while the triangles are for the fit for the 5-flavor scheme (scenario iii).}  
\label{fig:sigma}
\end{center}
\end{figure}

The quantitative description of the data with the statistical model is, however,
rather poor and certainly no improvement is visible for the case of subtracting 
charm and bottom contributions.
The poor fit quality which is already visible in Fig.~\ref{fig:91GeV} becomes
striking when, as in  Fig.~\ref{fig:sigma}, we show for the four energies 
the difference $\Delta$ (in units of the experimental error) between the 
experimental data and the statistical model calculations for the best fit values.

A summary of the fit parameters obtained for the various data sets is presented
in Table~\ref{tabl:param}. The errors of the thermal parameters listed in
Table~\ref{tabl:param} are the statistical errors as extracted from a 
minimization done with MINUIT \cite{minuit}, interfaced with our code. 
For the discussion of the systematic errors see below.

\begin{table}[htb]
   \caption{Values of fit parameters and $\chi^2$ values per degree of freedom 
   for different energies. At 91 GeV the fit results are given for the three
scenarios discussed in the text.} 
   \label{tabl:param}  
   \begin{tabular}{ccccc}
      \hline
      $\sqrt{s}$[GeV] & T[MeV] & V[fm$^3$] & $\gamma_s$ & $\chi^2$/dof\\
      \hline      
      10 & 166$\pm$1.7 & 10$\pm$1.5 & 0.80$\pm$0.02 & 314/20 \\
      29-35 & 162$\pm$1.7 & 16$\pm$1.4 & 0.96$\pm$0.03 & 108/17 \\
      91 i) & 160$\pm$0.5 & 26$\pm$1 & 0.82$\pm$0.007 & 595/28 \\
      91 ii) & 168$\pm$0.5 & 18$\pm$1 & 0.66$\pm$0.01 & 994/28 \\
      91 iii) & 170$\pm$0.5 & 16$\pm$1 & 0.66$\pm$0.01 & 499/28 \\
      130-200 & 154$\pm$2.8 & 40$\pm$4.3 & 0.82$\pm$0.03 & 12/2 \\
      \hline
   \end{tabular}
\end{table}

Typical $\chi^2$ values per degree of freedom lie between 5 and 20 and 
discrepancies between single data points and fit values larger than 5 standard 
deviations are not rare. Furthermore, there is no clear pattern observed: 
the fits are comparably poor for baryons and mesons, as well as for non-strange 
and strange hadrons.
In particular, for all energies the yields of $\phi$ mesons and of hyperons are 
poorly reproduced. Large deviations are seen also for kaons.

The case of the fit after subtraction of the contribution of the decays of 
charmed and bottom hadrons (scenario ii), explored at $\sqrt{s}$=91 GeV, 
is characterized by a larger $\chi^2$ value compared to the overall fit 
(scenario i).
The extracted fit parameters differ somewhat in the two cases.
In particular, T is higher and V is smaller for scenario ii) compared to 
scenario i), following the (T,V) correlation discussed 
below. Scenario iii) gives the lowest $\chi^2$ value, although the value
is still far from that of a good fit.

It should be mentioned that, while the general agreement among the four LEP
experiments is excellent, the measured yields of $\Sigma^*$ hyperons differ by
more than 70\%.  Excluding the $\Sigma^*$ hyperons would cause an increase of
$\gamma_s$ at $\sqrt{s}$=91 GeV, because the $\Sigma^*$ yields calculated in
the model are overestimated (see Fig. \ref{fig:91GeV}).  
This would slightly improve the situation for the $\Xi$ 
and $\Lambda$ multiplicities which are higher than those predicted by the 
model, but with only a marginal improvement of the $\chi^2$ values, as is 
discussed below.

\begin{table}
\caption{Calculated yields (corresponding to the fit of the subtracted data 
at 91 GeV, T=168 MeV, V=18 fm$^3$ and $\gamma_s$=0.66) for selected hadron species 
for the non-flavor scheme and for the 5-flavor one assuming vanishing or
fractional baryon and charge quantum numbers.
In the third column we show the percentage of the yields arising from the 
contribution of charm and bottom events, as calculated from the experimental data
(complemented by calculations using scenario iii) for $\Xi_c$, $\Omega_c$ and 
$\Omega_b$ hadrons).
}
\label{tab:comp1}
\begin{tabular}{cccc}
\hline
particle & Calculations & Data without & contribution\\
~ & ~ & $c$,$b$ contribution  & from $c$,$b$ (in \%)\\
\hline
$\pi^{+}$ & 15.80 & 14.97 & 12.0\\
$\pi^0$   & 9.45 & 8.50 & 9.8\\
K$^+$    & 1.43 & 1.69 & 24.2\\
$\rho^0$(770) & 1.042 & 1.209 & 1.8\\
K$^{*0}$(892) & 0.436 & 0.630 & 14.7\\
p           & 0.965 & 0.992 & 5.5\\
$\phi$(1020) & 0.090 & 0.0.054 & 44.2\\
$\Lambda$ & 0.300 & 0.336 & 14.3 \\
$\Sigma^{+}$(1385) & 0.0337 & 0.0232 & 3.0\\
$\Xi^-$ & 0.0123 & 0.0231 & 10.6 \\
$\Omega$ & 0.00056 & 0.00140 & 12.8 \\
\hline
\end{tabular}
\end{table}

In Table~\ref{tab:comp1} we show for selected hadron species the calculations 
in the no flavor scheme and the experimental data after the subtraction
of the charm and beauty contribution. The relative magnitude of this contribution
is also listed in Table~\ref{tab:comp1}.

\begin{figure}[htb]
\begin{tabular}{cc} 
\begin{minipage}{.49\textwidth}
\centering\includegraphics[width=1.03\textwidth]{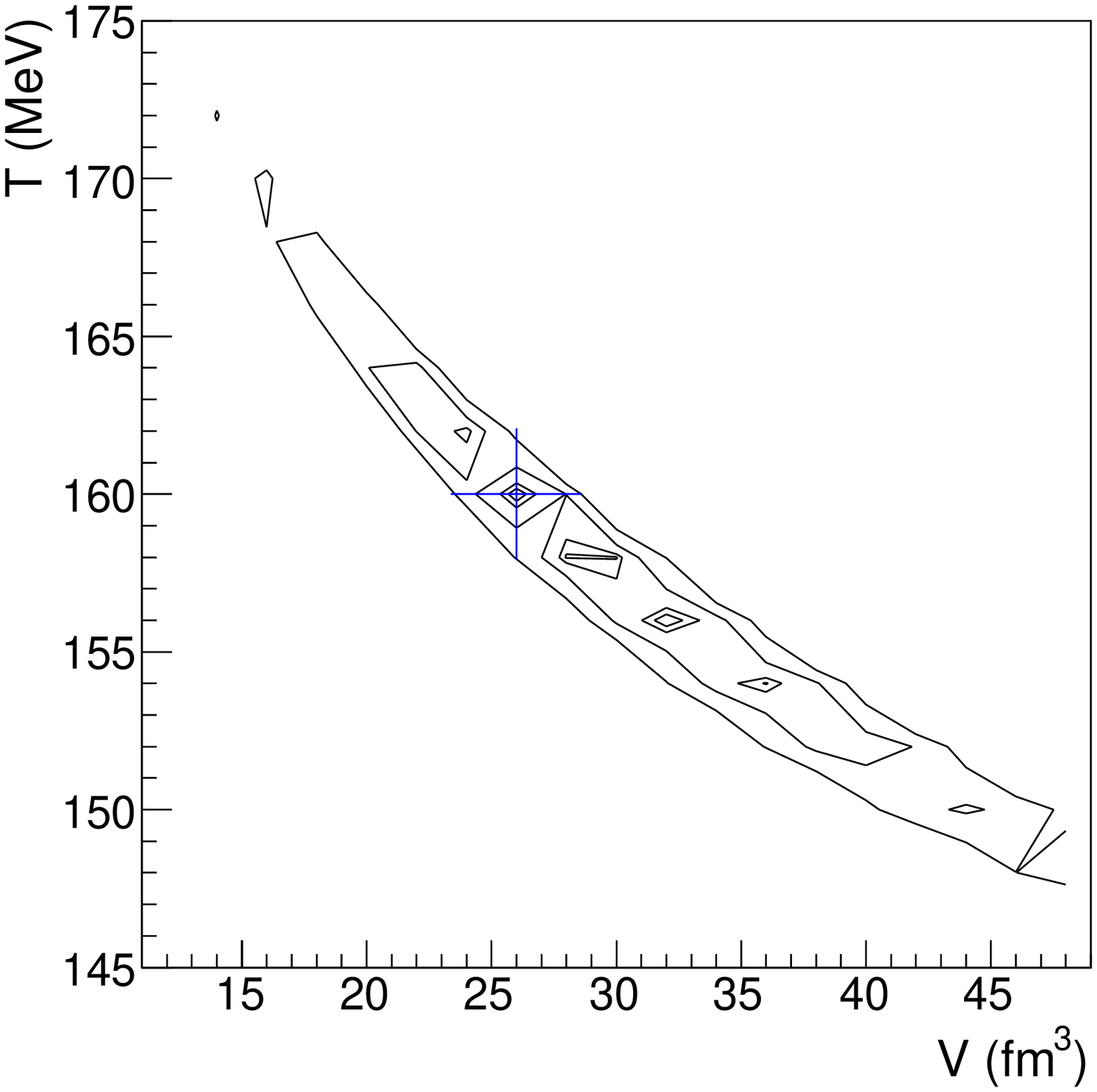}
\end{minipage} & \begin{minipage}{.49\textwidth}
\centering\includegraphics[width=1.03\textwidth]{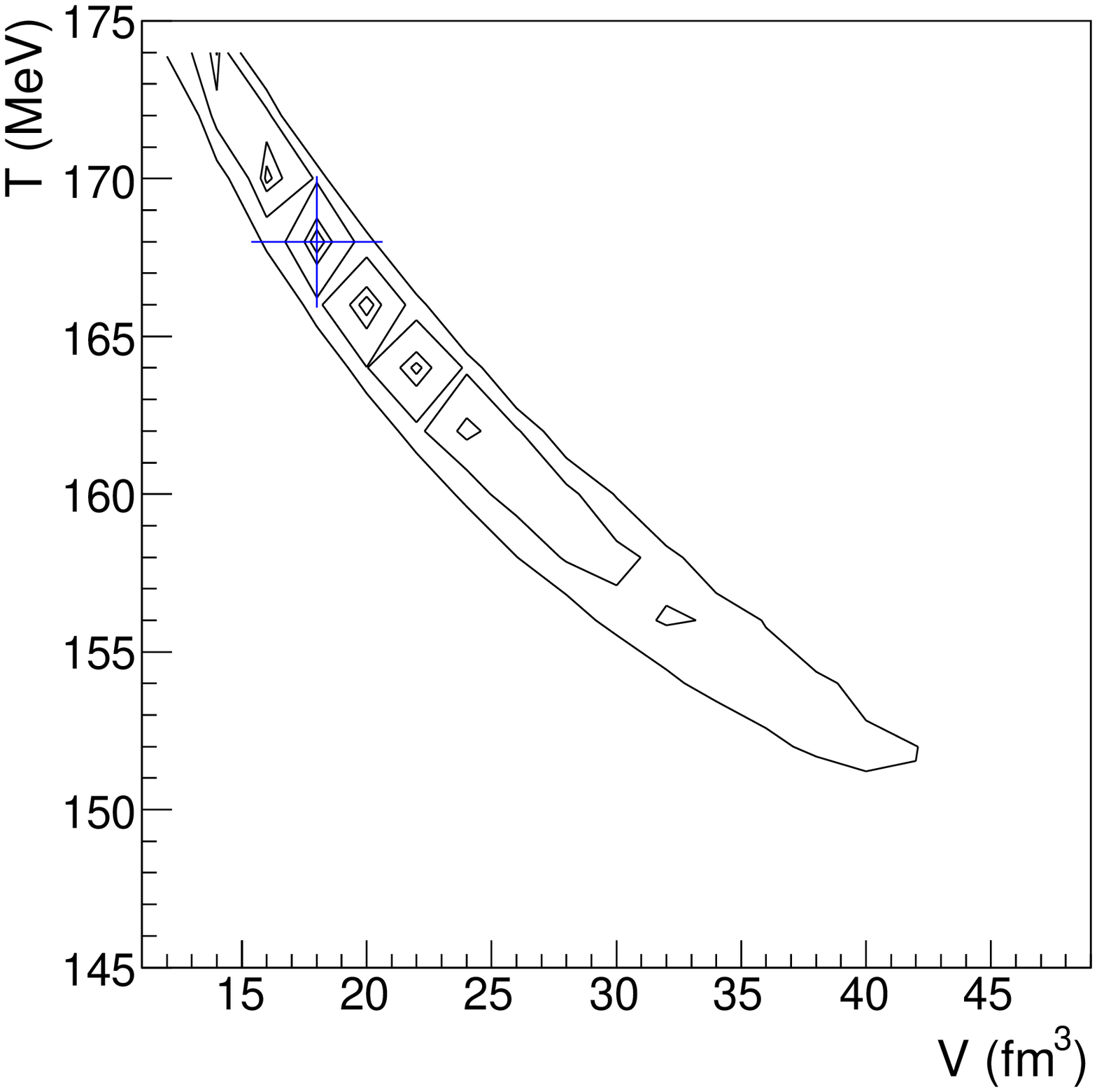}
\end{minipage}
\end{tabular}
\caption{$\chi^2$ contour lines in temperature and volume space for
the overall fit (left panel) and after subtraction of charm and bottom decays 
(right panel) for $\sqrt{s}$=91 GeV for the no flavor scheme. 
The contours correspond to $\chi^2_{min}$+10\%, +20\%, +50\%, +100\%. 
The best fit values are indicated by the crosses.}
\label{fig:91GeVcont}
\end{figure}

A difficulty for the determination of fit parameters is visible if one inspects 
the $\chi^2$ contour lines as shown in Fig.~\ref{fig:91GeVcont} in (T,V) plane
for fits at 91 GeV. 
One notices in this figure a strong anticorrelation between the fit parameters 
which is also present in the (T,$\gamma_s$) space (not shown).
Closer inspection reveals, in addition, a series of local minima which
indicates the difficulty in the determination of the fit parameters.  Such
local minima are typical for poor fits and imply that the true uncertainties 
in the fit parameters are likely much larger than the values obtained from 
the standard fit procedure \cite{minuit} employed here.

Despite these caveats about fit quality and uncertainties it is noteworthy
that the temperature parameters obtained from most data sets are close to 160
MeV and nearly independent of energy, similar to results of previous
investigations. In contrast, the volume increases with the center of mass
energy. 
The values obtained for the strangeness undersaturation parameter 
$\gamma_s$ range between 0.96 and 0.70 and exhibit no clear trend with energy.
For the 5-flavor case, at 91 GeV we obtain a volume of 18 fm$^3$. 
This is smaller than the results reported in \cite{becattini08} 
by about a factor of 2 because, in our case, the 2 jets hadronize separately.

In the following we will take a closer look at the 91 GeV fit, as the data set
at that energy contains the largest number of measured hadron yields. 
We consider scenario i), that is, the fit to data without subtracting the 
charm and bottom decay contributions to calculations in the non-flavor scheme.
To check whether the high $\chi^2$/dof values are caused by discrepancies 
for a few particular particles, we excluded 3 hadron species 
(the $\Sigma^*$'s and $\phi$) and repeated the fit. 
Naturally, the fit is better, but we still found a high $\chi^2$/dof=208/25 
(for T=158 MeV, V=28 fm$^3$, $\gamma_s$=0.90).

The data from LEP comprise very many particle species and their yields
typically are measured with accuracies of a few percent. Clearly 
the precision of the experimental data set provides a stringent test of the
statistical model. To provide a quantitative estimate at which accuracy level
the statistical model breaks down and to allow comparison with
the situation encountered for RHIC data we performed a fit at $\sqrt{s} = 91$
GeV to those hadron yields which were used for the fit of central
nucleus-nucleus collision data at $\sqrt{s_{NN}}$=200 GeV \cite{aat}, namely
$\pi^+$, $\pi^0$, $K^+$, $K^0$, $K^{*+}$, $K^{*0}$, $p$, $\Lambda$, $\Xi^-$, 
$\Omega^-$, and $\phi$. 
The resulting thermal parameters are: T=162$\pm$2 MeV V=24$\pm$2 fm$^3$, and 
$\gamma_s$=0.82$\pm$0.02, with a still poor $\chi^2$/dof of 278/8.
Excluding the $\phi$-meson from the fit yields T=160$\pm$2 MeV, V=24$\pm$2 fm$^3$ 
and $\gamma_s$=0.96$\pm$0.02, with  $\chi^2$/dof = 62/7.
A more direct comparison between thermal fits for heavy ion and $e^+e^-$ data 
can be obtained by arbitrarily assigning the uncertainties of the RHIC data 
to be corresponding LEP yields (namely uncertainties of the order of 10\%). 
Constraining, as for the RHIC data, the fit parameters to T and V, i.e. setting 
$\gamma_s = 1$  yields then T = 158$\pm$2 MeV and V = 24$\pm$2 fm$^3$, with 
$\chi^2$/dof = 49/9. 
A reasonable fit can only be obtained by also letting $\gamma_s$ vary freely, 
with resulting parameters T=168$\pm$2 MeV, V=18$\pm$2fm$^3$ and 
$\gamma_s$=0.80$\pm$0.02, with $\chi^2$/dof = 18/8. 
These exercises demonstrate that a statistical model description of $e^+e^-$ data 
fails badly without the introduction of the non-equilibrium parameter $\gamma_s$.
Even using $\gamma_s$ the statistical description breaks down completely at 
an accuracy level for the data better than 10\%. 

Another noteworthy difference between fireballs in $e^+e^-$ and
nucleus-nucleus collisions is their energy content. For its determination we
have computed the energy density $\epsilon$ in the hadronic gas to yield the 
thermal energy content E$_{j} = \epsilon V$ of the jet at chemical decoupling.
For the $e^+e^-$ case and the parameters reported in Table~\ref{tabl:param},
this procedure leads to E$_{j}$ = 3.23, 5.45, 8.6, 10.2 GeV at $\sqrt{s}$ = 10,
29-35, 91, 130-200 GeV, respectively.  Hence, the thermal energy within
each jet is only a small fraction of $\sqrt{s}/2$, 
(e.g. about 19\% at 91 GeV).  Apparently, in the thermal interpretation of 
$e^+e^-$ collisions, most of the c.m. energy is not available for particle 
production. This is in strong contrast with results for nucleus-nucleus collisions.
We have analyzed for this purpose central collision events for 20 and 40 
GeV/nucleon Pb-Pb collisions \cite{aat}, where it makes sense to consider data 
integrated over the full phase space. In these cases we find that the energy 
content in the fireball amounts to 61 \% and 63 \% of the total c.m. energy 
at 20 and 40 AGeV, respectively, implying that most of the total c.m. energy 
in a nucleus-nucleus collision is thermal, with the remaining non-thermal 
fraction likely to be due to collective flow.
We note that these  differences  are not consistent with the
finding that particle production in $e^+e^-$, pp and nucleus-nucleus
collisions is universally governed by the available c.m. energy
\cite{phobos,cley2,basile}.

\section{Summary and conclusions} 

We analyzed the comprehensive set of measured yields of hadrons with 
light quarks ($u$, $d$, $s$) in $e^+e^-$ collisions in a range from 10 GeV 
up to 200 GeV within an equilibrium thermodynamics picture. 
The calculations were performed in the framework of canonical partition 
functions to conserve explicitely baryon number, strangeness, charge,
charmness and bottomness and made use of the hadron resonance gas 
description. 
Our results corroborate previous findings 
\cite{Becattini:1995if,Becattini:1996gy,becattini01,Becattini:1997uf} 
that statistical features are present in hadron production in e$^+$e$^-$ 
collisions.
At $\sqrt{s}$=91 GeV, three scenarios were considered, namely fitting data
without and with subtraction of the decay products of charmed and bottom hadrons
in our no flavor approach (vanishing jet quantum numbers) and 
employing a 5-flavor scheme where the $q$ and $\bar{q}$ jets (with the 
relative abundancies of the $q\bar{q}$ events as extra parameters taken
from measurements at LEP) are carrying the flavor quantum numbers.
The two thermodynamical parameters temperature T and volume V,
and a strangeness undersaturation factor $\gamma_s$ were obtained from a
$\chi^2$ minimization procedure. 
While we find, as in previous investigations
\cite{Becattini:1995if,Becattini:1996gy,becattini01,Becattini:1997uf}, 
that the resulting temperature value is close to 160 MeV, independent of 
energy\footnote{In
  contrast, analysis of nuclear fireballs \cite{aat} yields temperature values
  which decrease with decreasing energy.}, the overall description of the 
high-precision LEP data is rather poor, independent of whether heavy quark 
contributions are subtracted or not.
The $\chi^2$/dof values larger than 5 for all fits call into
question the validity of the thermodynamical approach for these data. 
This conclusion still holds even if the analysis is restricted to the same set
of hadrons which were analyzed in the context of thermal model fits to Au-Au
collision data from the RHIC accelerator.

The apparent statistical fingerprint visible in the LEP data and first
observed in \cite{Becattini:1995if} breaks down at an accuracy level of about
10\%. Even at that level the $e^+e^-$ data cannot at all be described without
the explicit assumption of strangeness undersaturation, implying that hadron
production in $e^+e^-$ originates from a state which is quite far from true
thermodynamic equilibrium. This conclusion is  further supported by the
observation that the corresponding fireball volume contains only a small
fraction of the overall c.m. energy, implying that most of the c.m. energy 
is not available for particle production.
This is in strong contrast to the situation in nucleus-nucleus collisions.

A striking feature observed for nuclear fireballs is the complete absence of
strangeness suppression. In this context it would be very interesting to probe
the equilibrium features observed in central nuclear collisions with increasing
accuracy. Will non-equilibrium features there also be first revealed in the
strangeness sector?  To clarify these issues will be a challenge for future
measurements.

\section*{Acknowledgments}
We acknowledge illuminating discussions with Francesco Becattini.
We acknowledge the support of the Alliance Program of the Helmholtz 
Association HA216/EMMI.
K.R. acknowledges partial support from the Polish Ministry of Science (MENiSW)
and the Deutsche Forschungsgemeinschaft (DFG) under the Mercator Programme.
F.B acknowledges helpful discussions with Kai Schweda.


\begin{thebibliography}{99}


\bibitem{agssps} P. Braun-Munzinger, J. Stachel, J.P. Wessels and
  N. Xu, Phys. Lett. B {\bf 344} (1995) 43 [nucl-th/9410026] and 
Phys. Lett. B {\bf 365} (1996) 1 [nucl-th/9508020].

\bibitem{satz} J. Cleymans, D. Elliott, H. Satz, and R.L. Thews, Z. Phys. C
{\bf 74} (1997) 319 [nucl-th/9603004].

\bibitem{heppe} P. Braun-Munzinger, I. Heppe and J. Stachel, Phys. Lett. B
{\bf 465} (1999) 15 [nucl-th/9903010].

\bibitem{cley} J. Cleymans and K. Redlich, Phys. Rev. C {\bf 60} (1999) 054908
[nucl-th/9903063].

\bibitem{beca1} F. Becattini, J. Cleymans, A. Keranen, E. Suhonen, and
K. Redlich, Phys. Rev. C {\bf 64} (2001) 024901 [hep-ph/0002267].

\bibitem{rhic} P. Braun-Munzinger, D. Magestro, K. Redlich, and
  J. Stachel, Phys. Lett. B {\bf 518} (2001) 41 [hep-ph/0105229].

\bibitem{nu} N. Xu and M. Kaneta, Nucl. Phys. A {\bf 698} (2002) 306c.

\bibitem{beca2} F. Becattini, J. Phys. G {\bf 28} (2002) 1553.

\bibitem{rapp} R. Rapp and E. Shuryak, Phys. Rev. Lett. {\bf 86} (2001) 2980
[hep-ph/0008326].

\bibitem{becgaz} F. Becattini, M. Ga\'zdzicki, J. Manninen, Phys. Rev. C {\bf 73}
(2006) 044905 [hep-ph/0511092].

\bibitem{aat} A. Andronic, P. Braun-Munzinger, J. Stachel, 
Nucl. Phys. A {\bf 772} (2006) 167 [nucl-th/0511071].

\bibitem{review}P. Braun-Munzinger, K. Redlich, and J. Stachel,
  nucl-th/0304013, invited review in Quark Gluon Plasma 3, eds. R.C. Hwa and
  X.N. Wang, (World Scientific Publishing, 2004).
 
\bibitem{Becattini:1995if}
  F.~Becattini,
  Z. Phys. C {\bf 69} (1996) 485.


\bibitem{Becattini:1996gy}
  F.~Becattini,
  hep-ph/9701275.

\bibitem{becattini01}
  F.~Becattini, G.~Passaleva, Eur. Phys. J. C {\bf 23} (2002) 551 
[hep-ph/0110312].

\bibitem{Becattini:1997uf}
  F.~Becattini,
  J.\ Phys.\ G {\bf 23} (1997) 1933
  [hep-ph/9708248].

\bibitem{stock} R. Stock, Phys. Lett. B {\bf 465} (1999) 277
[hep-ph/9905247].

\bibitem{heinz} U. Heinz, Nucl. Phys. A {\bf 685} (2001) 414 
[hep-ph/0009170].

\bibitem{castorina} P. Castorina, D. Kharzeev, H. Satz, 
Eur. Phys. J. C {\bf 52} (2007) 187 [arXiv:0704.1426].

\bibitem{wetterich} P. Braun-Munzinger, J. Stachel, C. Wetterich,
  Phys. Lett. B {\bf 596} (2004) 61 [nucl-th/0311005].

\bibitem{pdg}
C. Amsler et al. [Particle Data Group], Phys. Lett. B 667 (2008) 1.

\bibitem{redlich} R. Hagedorn, K. Redlich, Z. Phys. C {\bf 27} (1985) 541.

\bibitem{becattini08}
F. Becattini, P. Castorina, J. Manninen, H. Satz,  
Eur. Phys. J. C {\bf 56} (2008) 493 [arXiv:0805.0964].

\bibitem{xyz} A. Andronic et al., arXiv:0904.1368 [hep-ph].

\bibitem{turko81} L. Turko, Phys. Lett. B {\bf 104} (1981) 153.

\bibitem{beutler_diploma} F. Beutler, diploma thesis, Univ. of Heidelberg, 2008
and paper in preparation.

\bibitem{aas} A. Andronic, P. Braun-Munzinger, J. Stachel, 
Phys. Lett. B {\bf 673} (2009) 142 [arXiv:0812.1186].

\bibitem{cle91} 
J. Cleymans, K. Redlich, E. Suhonen, Z. Phys. C {\bf 51} (1991) 137.

\bibitem{Cleymans:1997ib}
  J.~Cleymans, M.~Marais and E.~Suhonen,
  Phys.\ Rev.\  C {\bf 56} (1997) 2747
  [nucl-th/9705014].

\bibitem{pbm4} P. Braun-Munzinger, J. Cleymans, H. Oeschler, K. Redlich,  
Nucl. Phys. A {\bf 697} (2002) 902 [hep-ph/0106066].

\bibitem{Wheaton:2004qb}
  S.~Wheaton and J.~Cleymans,
  hep-ph/0407174; J. Phys. G {\bf 31} (2005) S1069.

\bibitem{minuit} MINUIT, Function Minimization and Error Analysis, CERN 
Program Library Long Writeup D506, http://wwwasdoc.web.cern.ch/wwwasdoc/minuit/

\bibitem{phobos} B.B. Back et al., PHOBOS coll., nucl-ex/0301017.

\bibitem{cley2} J. Cleymans, M. Stankiewicz, P. Steinberg, S. Wheaton, 
nucl-th/0506027.

\bibitem{basile} M. Basile et al., Phys. Lett. B {\bf 92} (1980) 367,
  Phys. Lett. B {\bf 95} (1980) 311. 

\end{thebibliography}
\end{document}